\documentclass[conference]{IEEEtran}
\IEEEoverridecommandlockouts

\usepackage{cite}
\usepackage{amsmath,amssymb,amsfonts}
\usepackage{algorithmic}
\usepackage{graphicx}
\usepackage{textcomp}
\usepackage{xcolor}
\usepackage{bm} 
\usepackage{float}
\usepackage{caption}
\usepackage{multirow}
\usepackage{pict2e} 
\usepackage{array} 
\usepackage{pict2e} 
\usepackage{boldline} 
\usepackage{booktabs} 
\usepackage{tabularx}
\usepackage{marvosym}
\usepackage{siunitx}
\usepackage[colorlinks=true, linkcolor=black, citecolor=black, urlcolor=black]{hyperref}
\usepackage[numbers,sort&compress]{natbib}
\usepackage{subcaption}
\usepackage{url}
\usepackage[utf8]{inputenc}

\newcommand{\ie}[0]{\textit{i.e.}}
\def\BibTeX{{\rm B\kern-.05em{\sc i\kern-.025em b}\kern-.08em
    T\kern-.1667em\lower.7ex\hbox{E}\kern-.125emX}}
\begin{document}

\title{Multi-Prototype Embedding Refinement for Semi-Supervised Medical Image Segmentation
\thanks{}
}

\author{\IEEEauthorblockN{Yali Bi$^1$\IEEEauthorrefmark{1}, Enyu Che$^1$\IEEEauthorrefmark{1}, Yinan Chen$^1$, Yuanpeng He$^2$, and Jingwei Qu$^1$\IEEEauthorrefmark{2}}
	\IEEEauthorblockA{$^1$College of Computer and Information Science, Southwest University, Chongqing, P.R. China\\
		$^2$School of Computer Science, Peking University, Beijing, P.R. China\\
	}
}

\maketitle

\begin{abstract}
Medical image segmentation aims to identify anatomical structures at the voxel-level. Segmentation accuracy relies on distinguishing voxel differences. Compared to advancements achieved in studies of the inter-class variance, the intra-class variance receives less attention. Moreover, traditional linear classifiers, limited by a single learnable weight per class, struggle to capture this finer distinction. To address the above challenges, we propose a Multi-Prototype-based Embedding Refinement method for semi-supervised medical image segmentation. Specifically, we design a multi-prototype-based classification strategy, rethinking the segmentation from the perspective of structural relationships between voxel embeddings. The intra-class variations are explored by clustering voxels along the distribution of multiple prototypes in each class. Next, we introduce a consistency constraint to alleviate the limitation of linear classifiers. This constraint integrates different classification granularities from a linear classifier and the proposed prototype-based classifier. In the thorough evaluation on two popular benchmarks, our method achieves superior performance compared with state-of-the-art methods. Code is available at \url{https://github.com/Briley-byl123/MPER}.
\end{abstract}


\section{Introduction}
\label{sec:intro}

Medical image segmentation aims to accurately identify anatomical structures like organs and tumors~\cite{ref1}, and has been utilized in various fields, such as disease diagnosis~\cite{ghoshal2021estimating, huang2025unitrans}. It is more challenging than natural image segmentation since the data labeling requires medical expertise. Moreover, the inherent uncertainties in medical images—such as ambiguous boundaries, noise, and inter-observer variability—further complicate the segmentation task. Addressing these uncertainties effectively is crucial for improving model reliability and robustness \cite{he2024generalized}.

Semi-supervised frameworks are increasingly applied in medical image segmentation due to their advantages in exploiting unlabeled data. These methods explore different optimization strategies for predictions of unlabeled data, including consistency regularization~\cite{ref4,he2024mutual}, entropy minimization~\cite{ref19,ref20}, and pseudo-labeling~\cite{ref13,KongRL2024Semi}. Nonetheless, the progress in studying the intra-class voxel variance remains limited. Additionally, existing methods often lack a principled way to handle uncertainty, which can lead to unreliable predictions in ambiguous regions.

Uncertainty estimation has been widely studied to quantify model confidence and improve decision-making in segmentation tasks. In this context, evidence theory (Dempster-Shafer theory) provides a robust framework for modeling uncertainty by associating evidence with different hypotheses \cite{he2021conflicting, he2022mmget, he2022new}. Unlike probability theory, which assigns precise probabilities, evidence theory allows the integration of multiple sources of information, capturing both aleatoric uncertainty (stemming from data variability) and epistemic uncertainty (due to model limitations) \cite{he2023ordinal, he2023tdqmf, he2022ordinal}. Applying evidence theory to medical image segmentation can enhance the model’s ability to deal with ambiguous regions and refine predictions based on confidence-aware decision-making. However, integrating evidence theory into deep learning-based segmentation frameworks remains underexplored \cite{he2024residual}.

\begin{figure}[!t]
	\centering
	\includegraphics[width=0.95\linewidth]{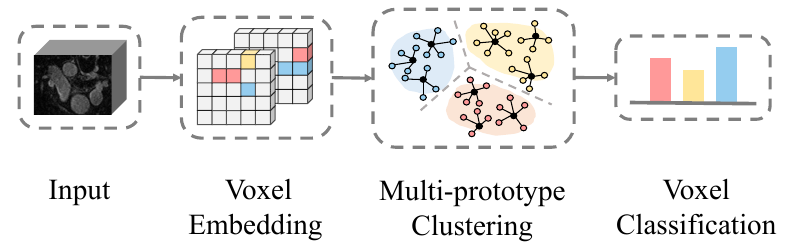}
	\caption{Multi-prototype-based classification. Voxel embeddings extracted from input medical images are assigned to the most similar prototypes. The resulting clusters help generate more precise segmentation through finer classification.}
	\label{fig:coremind}
	\vspace{-3mm}
\end{figure}

The quality of segmentation highly depends on capturing variations between voxels, as investigated in previous studies \cite{he2024uncertainty, he2024epl}. However, these studies focus on the inter-class variance, overlooking the finer distinctions between voxels belonging to the same class. In fact, the intra-class variance offers a more granular perspective for precise segmentation. Moreover, traditional linear classifiers learn a single weight for each class. The class-level weights have limited capacity to distinguish semantic ambiguities between voxels of the same class, thereby also hindering deeper exploration of the intra-class variance \cite{li2024efficient}. To address the above challenges, we propose a **Multi-Prototype-based Embedding Refinement** method for medical image segmentation. 

Prototype learning selects representative samples to summarize critical features and inherent structures of original data. It has been introduced into semi-supervised segmentation frameworks~\cite{ref26,Yuchao2022Semi}. A single prototype is generated for each object class via masked average pooling over voxel embeddings. However, a single prototype struggles to describe the inner variations in a class. Our solution incorporates a multi-prototype-based classification strategy (Fig.~\ref{fig:coremind}) into a semi-supervised framework, which rethinks medical image segmentation from the perspective of structural relationships between voxel embeddings. Multiple prototypes form several sub-centers of a class in the voxel embedding space. Their spatial distribution depicts inner structures of the class. Voxels cluster along this distribution based on their feature similarity with prototypes. This cluster approach implicitly reveals the structural relationships among voxels, simultaneously capturing their intra-class variations. To alleviate the limitation of linear classifiers \cite{li2025adaptive, xu2023spatio, li2022nndf}, we design a consistency constraint between a linear classifier and the proposed prototype-based classifier, enhancing segmentation quality by combining their classification granularities.

To further improve segmentation robustness, we integrate uncertainty modeling within our framework. Specifically, we leverage evidence theory to refine prototype-based classification by assigning confidence scores to voxel predictions. Each voxel is associated with multiple prototypes, and the model dynamically assigns belief masses to different segmentation hypotheses, mitigating the effect of ambiguous regions. This evidence-based refinement enhances segmentation performance, particularly in cases where voxel-level uncertainty is high.

Moreover, a strong **visual backbone** is essential for learning effective voxel embeddings. Modern deep learning-based segmentation models commonly adopt convolutional neural networks (CNNs) or vision transformers as their backbone architectures. While CNNs excel at capturing local spatial structures, vision transformers demonstrate superior global context modeling capabilities, which are particularly beneficial for medical image segmentation. In our approach, we adopt a hybrid backbone that leverages both CNN-based feature extraction and transformer-based context modeling, ensuring rich multi-scale representations. This hybrid strategy strengthens the learned voxel embeddings, enabling our multi-prototype classification scheme to better capture intra-class variations.

Experiments on two benchmarks demonstrate our method's superior performance, both quantitatively and qualitatively, compared to state-of-the-art (SOTA) methods. Our findings highlight the importance of modeling intra-class variance, incorporating uncertainty-aware learning, and leveraging advanced visual backbones to achieve more accurate and reliable medical image segmentation.

\begin{figure*}[!t]
    \centering
    \includegraphics[width=0.8\linewidth]{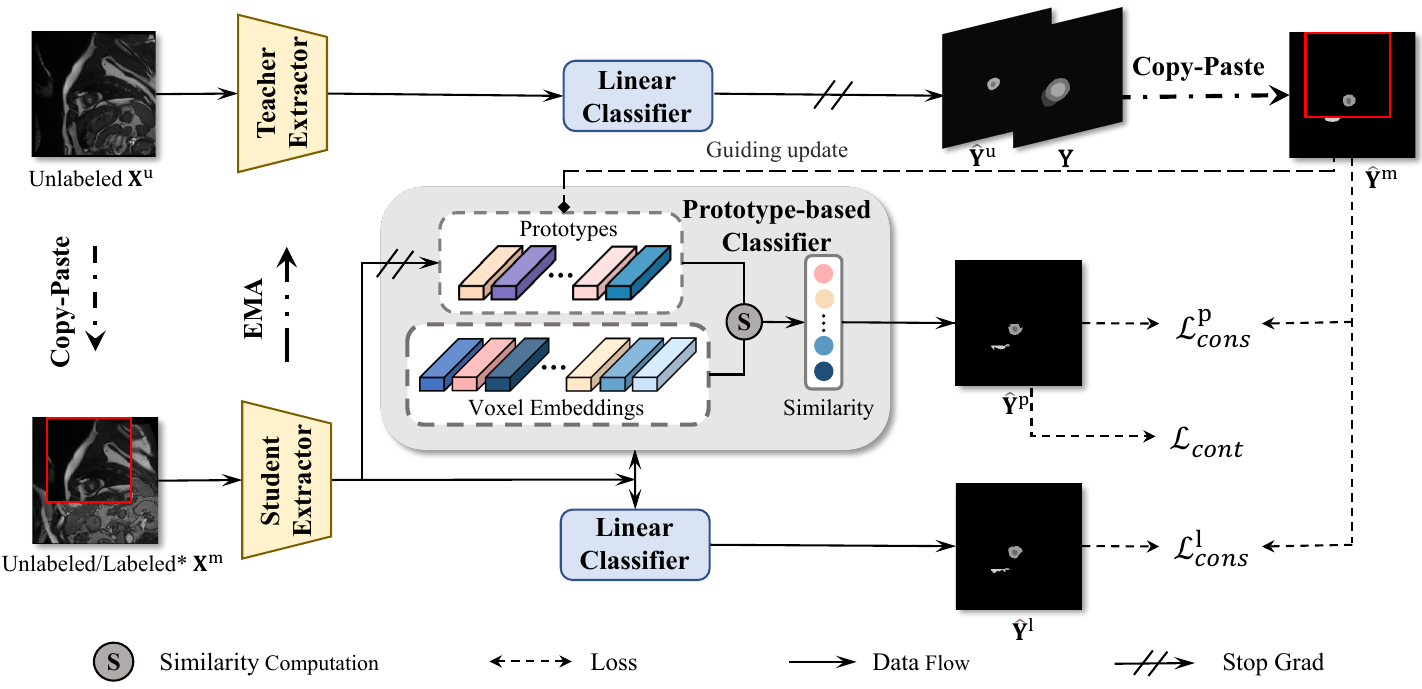}
    \caption{Overview of our method.}
    \label{fig:overview}
    \vspace{-3mm}
\end{figure*}


\section{Methodology}
\label{sec:method}
We implement our method based on the classical Mean Teacher model~\cite{tarvainen2017mean} (Fig.~\ref{fig:overview}). Our model is trained by a two-stage strategy: pre-training and self-training. During the pre-training, we utilize labeled images to train a supervised feature extractor. The Copy-Paste augmentation~\cite{ref37} is adopted to enrich the training data. One image is randomly cropped and pasted onto another to obtain a mixed image. 

In the self-training stage, we introduce the multi-prototype-based classification strategy and the consistency constraint. The extractors of the Teacher and Student networks are both initialized using the pre-trained feature extractor. The Teacher network processes an unlabeled image $\mathbf{X}^\mathrm{u} \in \mathbb{R}^{H \times W \times D}$ to generate a pseudo-label $\hat{\mathbf{Y}}^\mathrm{u} \in \mathbb{R}^{H \times W \times D}$, with weights updated through Exponential Moving Average (EMA) of the Student network. $W$, $H$, and $D$ denote the width, height, and depth of the image, respectively. The Student network receives a mixed image $\mathbf{X}^\mathrm{m}$ generated by bidirectional Copy-Paste augmentation~\cite{ref27}, where the unlabeled image $\mathbf{X}^\mathrm{u}$ is randomly cropped and pasted onto a labeled image $\mathbf{X}^\mathrm{l}$, or vice versa. For clarity, only the former case is shown in Fig.~\ref{fig:overview}. This augmentation strategy enables the semi-supervised model to better incorporate information from unlabeled data. The embedding $\mathbf{Z} \in \mathbb{R}^{H \times W \times D \times d}$ of the input image $\mathbf{X}^\mathrm{m}$ is obtained by the Student extractor, where $d$ denotes the embedding dimensions. Then, the prototype-based classifier predicts the label $\hat{\mathbf{Y}}^\mathrm{p}$ based on the embedding $\mathbf{Z}$. Analogously, another label $\hat{\mathbf{Y}}^\mathrm{l}$ is estimated by a linear classifier. In addition, we combine the ground-truth label $\mathbf{Y} \in \left\{0, 1, \dots, C - 1\right\}^{H \times W \times D}$ of the labeled image $\mathbf{X}^\mathrm{l}$ and the pseudo-label $\hat{\mathbf{Y}}^\mathrm{u}$ to obtain a mixed label $\hat{\mathbf{Y}}^\mathrm{m}$, where $C$ is the number of object classes. This is done similarly to how the input $\mathbf{X}^\mathrm{m}$ is processed. $\hat{\mathbf{Y}}^\mathrm{m}$ provides the supervision information for the Student network.

\subsection{Multi-Prototype-based Classification}
\label{sec:method-1}
A prototype is defined as a representative of a class in the embedding space. Built upon this, multiple prototypes of the class form different sub-centers, and implicitly divide the embedding space at a finer granularity level. Formally, we define $K$ prototypes for each class. The $k$-th prototype of the $c$-th class is denoted as $\mathbf{p}_c^k \in \mathbb{R}^d$, where $c \in \left\{0, 1, \dots, C - 1\right\}$ and $k \in \left\{0, 1, \dots, K - 1\right\}$.

\noindent\textbf{Prototype-based Classifier.}\ To estimate the category of each voxel, we compute its posterior probability distribution by measuring its similarity with all prototypes. Specifically, the cosine similarities between the voxel embedding $\mathbf{z} \in \mathbb{R}^d$ and all prototypes $\left\{\mathbf{p}_c^k\right\}_{c,k=0}^{C - 1, K - 1}$ are computed. Next, the highest similarity with $K$ prototypes of each class forms a similarity vector, which is further transformed into the probability distribution $\hat{\mathbf{y}}^\mathrm{p} \in \mathbb{R}^C$ by the softmax function:
\vspace{-1.5mm}
\begin{equation}
\vspace{-0.5mm}
    \hat{\mathbf{y}}^\mathrm{p}_c = \frac{\exp\left(\text{sim}\left(\mathbf{z}, \mathbf{p}_c^{k^\ast}\right)/ \tau_1\right)}{\sum_{c^\prime=0}^{C-1}\exp\left(\text{sim}\left(\mathbf{z}, \mathbf{p}_{c^\prime}^{k^\ast}\right)/ \tau_1 \right)}
\end{equation}
 where $\hat{\mathbf{y}}^\mathrm{p}_c$ denotes the probability that the voxel belongs to the $c$-th class, $\text{sim}(\cdot, \cdot)$ is defined as the cosine similarity function between two vectors, $k^\ast$ is the index of the prototype with the highest similarity in each category, and $\tau_1$ is a temperature parameter~\cite{wu2018unsupervised}. The class to which the prototype with the highest similarity belongs is the predicted class of the voxel.
 
\noindent\textbf{Prototype Initialization and Update.}\ Prototypes are initialized as the clustering results of the voxel embeddings through the MiniBatchKMeans algorithm~\cite{sculley2010web}. To update prototypes, we design a momentum-based strategy under the guidance of the mixed label generated from the Teacher network. First, the voxels of the input image $\mathbf{X}^\mathrm{m}$ are classified according to the mixed label $\hat{\mathbf{Y}}^\mathrm{m}$. Second, the voxels in each class further cluster into different groups based on their similarities with the multiple prototypes, \ie, each voxel is assigned to the prototype with the highest similarity. Finally, the embeddings of voxels in each cluster are utilized to update the prototype by the momentum-based strategy:
\vspace{-1.5mm}
\begin{equation}
\vspace{-0.5mm}
   \mathbf{p}_{t+1}=\eta\mathbf{p}_t + \left(1-\eta\right)\bar{\mathbf{z}}_t
\end{equation}
where $\bar{\mathbf{z}}_t$ denotes a mean embedding of the cluster at the $t$-th iteration, and $\eta \in \left(0, 1\right)$ is a momentum coefficient controlling the prototype update rate. To compute the mean embedding, we only keep the embeddings of unlabeled voxels whose pseudo-label confidence exceeds a given threshold $\alpha$. These are then used together with the embeddings of labeled voxels to compute the mean embedding. This ensures reliable embeddings are used, effectively integrating information from both labeled and unlabeled data.

\subsection{Consistency Constraint \& Training Objective}
\label{sec:method-2}
The traditional linear classifier shows robustness against noise and data incompleteness during the early training phases, thus enhancing training stability and accelerating model convergence, as investigated in previous studies. Specifically, it predicts the probability distribution $\hat{\mathbf{y}}^\mathrm{l}$ of each voxel by a linear layer with the softmax function:
\vspace{-1.5mm}
\begin{equation}
\vspace{-0.5mm}
    \hat{\mathbf{y}}^\mathrm{l}_c = \frac{\exp\left(\mathbf{w}_c^\top\mathbf{z}\right)}{\sum_{c^\prime=0}^{C-1}\exp\left(\mathbf{w}_{c^\prime}^\top\mathbf{z}\right)}
\end{equation}
where $\hat{\mathbf{y}}^\mathrm{l}_c$ is the probability that the voxel belongs to the $c$-th class, and $\mathbf{w}_c \in \mathbb{R}^d$ is the learnable weight of the $c$-th class.

\noindent\textbf{Consistency Constraint.}\ However, the class-level weights limit deeper exploration of the intra-class variance. Therefore, we build a consistency constraint to integrate different classification granularities from the linear classifier and the prototype-based classifier. The constraint is established by two consistency losses used for training our model: linear consistency loss $\mathcal{L}_{cons}^\mathrm{l}$ and prototype consistency loss $\mathcal{L}_{cons}^\mathrm{p}$. They are both implemented by Cross-Entropy loss $\mathcal{L}_{ce}$~\cite{zhang2018generalized} and Dice loss $\mathcal{L}_{dice}$~\cite{milletari2016v}:
\vspace{-1.5mm}
\begin{equation}
\vspace{-0.5mm}
    \mathcal{L}_{cons}^\mathrm{*} = \mathcal{L}_{ce}\left(\hat{\mathbf{Y}}^\mathrm{*}, \hat{\mathbf{Y}}^\mathrm{m}\right) + \mathcal{L}_{dice}\left(\hat{\mathbf{Y}}^\mathrm{*}, \hat{\mathbf{Y}}^\mathrm{m}\right)
\end{equation}
where $*$ denotes the superscript $\mathrm{l}$ or $\mathrm{p}$. The mixed label $\hat{\mathbf{Y}}^\mathrm{m}$ from the Teacher network acts as a bridge, aligning the prediction results $\hat{\mathbf{Y}}^\mathrm{p}$ and $\hat{\mathbf{Y}}^\mathrm{l}$ of the two classifiers. 

To further optimize the matching relationship between voxels and prototypes, we design a contrastive loss $\mathcal{L}_{cont}$ based on InfoNCE~\cite{Aaron2018CoRR}. Each voxel embedding $\mathbf{z}$ is treated as a query vector, and all prototypes $\left\{\mathbf{p}_c^k\right\}_{c,k=0}^{C - 1, K - 1}$ are viewed as key samples. It is assumed that the embedding $\mathbf{z}$ and its closest prototype $\mathbf{p}^\ast$ form a positive pair, with the remaining prototypes considered as negative samples for $\mathbf{z}$:
\vspace{-1.5mm}
\begin{equation}
\vspace{-0.5mm}
    \mathcal{L}_{cont} = -\log \frac{\exp\left(\mathbf{z}^\top\mathbf{p}^\ast / \tau_2\right)}{\sum_{c,k=0}^{C - 1, K - 1} \exp\left(\mathbf{z}^\top\mathbf{p}_c^k / \tau_2\right)}
\end{equation}
where $\tau_2$ denotes a temperature parameter that adjusts the sensitivity of the softmax function. 

Finally, we combine the three losses nonlinearly to guide voxels to the expected classes:
\vspace{-1.5mm}
\begin{equation}
\vspace{-0.5mm}
    \mathcal{L} = \mathcal{L}_{cons}^\mathrm{l} + \lambda(t)\left(\mathcal{L}_{cons}^\mathrm{p} + \gamma\mathcal{L}_{cont}\right)
\end{equation}
where $\gamma$ is a scaling factor controlling the relative importance of the contrastive loss $\mathcal{L}_{cont}$. $\lambda(t)$ is a dynamic weighting factor governed by a Sigmoid Ramp-up function, increasing the weight of prototype-associated losses from 0 to 1 over the training period. This dynamic weighting strategy ensures initial training stability through linear predictions while progressively enhancing the role of prototype-based predictions.

\begin{table}[!t]
    \caption{Comparison of segmentation quality on LA.}
    \label{tab:LA}
    \centering
    \sisetup{
		group-separator={,}, 
		group-four-digits=true, 
		table-align-text-post=false,
		detect-weight=true, 
		detect-family=true,
        mode=text
	}
        \begin{tabular}
        {r | l@{\hspace{1mm}} c | c@{\hspace{1mm}} c@{\hspace{1mm}} S[table-format=2.2]@{\hspace{1mm}} c}
        \toprule
        Method & Labeled & Unlabeled & Dice$\uparrow$ & Jaccard$\uparrow$ & {95HD$\downarrow$} & ASD$\downarrow$ \\
        \midrule
         & 4(5\%) & 0 & 52.55 & 39.60 & 47.05 & 9.87 \\
        V-Net~\cite{milletari2016v} & 8(10\%) & 0 & 82.74 & 71.72 & 13.35 & 3.26 \\
         & 80(All) & 0 & 91.47 & 84.36 & 5.48 & 1.51 \\
        \midrule
        UA-MT~\cite{ref30} & \multirow{9}*{4(5\%)} &\multirow{9}*{76(95\%)} & 82.26 & 70.98 & 13.71 & 3.82 \\
        SASSNet~\cite{ref31}& & & 81.60 & 69.63 & 16.16 & 3.58 \\
        DTC~\cite{ref32} &   &   & 81.25 & 69.33 & 14.90 & 3.99 \\
        URPC~\cite{ref34}&  &   & 82.48 & 71.35 & 14.65 & 3.65 \\
        MC-Net~\cite{ref33} &  &   & 83.59 & 72.36 & 14.07 & 2.70 \\
        SS-Net~\cite{ref29} &  &   & 86.33 & 76.15 & 9.97 & 2.31 \\
        MCF~\cite{ref36}  &  &   &82.56  &71.19  &16.05  &4.97  \\
        BCP~\cite{ref27} &  &   & 88.02& 78.72 & 7.90 & 2.15\\
        Ours&  &   &\textbf{88.82}& \textbf{80.00} & \hspace{1.7mm}\textbf{7.40} & \textbf{2.02}  \\
        \midrule
        UA-MT~\cite{ref30} & \multirow{9}*{8(10\%)} &\multirow{9}*{72(90\%)} & 87.79 & 78.39 & 8.68 & 2.12 \\
        SASSNet~\cite{ref31}  &  &  & 87.54 & 78.05 & 9.84 & 2.59 \\
        DTC~\cite{ref32}  &  &  & 87.51 & 78.17 & 8.23 & 2.36 \\
        URPC~\cite{ref34} &  &  &86.92 & 77.03 & 11.13 & 2.28 \\
        MC-Net~\cite{ref33} &  &  & 87.62 & 78.25 & 10.03 & 1.82 \\
        SS-Net~\cite{ref29} &  &  &88.55 & 79.62 & 7.49 & 1.90 \\
        MCF~\cite{ref36}   &    &   &88.71&80.41 &6.32 &1.90  \\
        BCP~\cite{ref27}&  &   &89.62 & 81.31 & 6.81& 1.76 \\
        Ours&  &   & \textbf{90.01}& \textbf{81.94}& \hspace{1.7mm}\textbf{6.51}& \textbf{1.74}\\
        \bottomrule
      \end{tabular}
  \vspace{-3mm}
\end{table}
\section{Experiments}
\label{sec:exp}
\subsection{Experimental Settings \& Implementation Details}
We evaluate our method against several SOTA methods on two popular benchmarks: LA~\cite{xiong2021global} and ACDC~\cite{bernard2018deep}. Following the general evaluation protocol~\cite{ref29}, we adopt 3D V-Net~\cite{milletari2016v} and 2D U-Net~\cite{Ronneberger2015U-net} as the feature extractors for the two datasets, respectively. For the LA dataset, training data consists of randomly cropped image patches of size $112 \times 112 \times 80$, with the zero-value region of mask defined as $74 \times 74 \times 53$, and data augmentation techniques (e.g., rotations, flipping) are applied. For the ACDC dataset, the randomly cropped zero-value area of the mask is defined as $170 \times 170$. For all experiments, we set $\eta = 0.999$, $\alpha = 0.9$, $\tau_1 = \tau_2 = 0.1$, and $K = 3$. Other hyperparameters are dataset-specific: (1) the scaling factor $\gamma = 0.02, 0.1$; (2) the batch size being 8 and 24; (3) the pre-training stage including 2k and 10k iterations; (4) the self-training stage including 15k and 30k iterations. All experiments are run on an NVIDIA GeForce RTX 4090 GPU.

\subsection{Evaluation Results}
\noindent\textbf{Quantitative Results.}\ The quantitative results in Tabs.~\ref{tab:LA} and~\ref{tab:ACDC} show that our method surpasses SOTA methods on both benchmarks. For the LA dataset, our method consistently leads across all metrics, especially with limited labeled data. At 5\% labeling, it achieves Dice coefficient of 88.82\% and Jaccard index of 80.0\%, with the lowest 95HD (7.4) and ASD (2.02) values. Similarly, for the ACDC dataset, our method performs strongly across varying amounts of labeled data. These results suggest that our method effectively utilizes unlabeled data. Notably, with only 10\% of annotations, our method shows a minimal Dice coefficient reduction of just 1.75\% compared to a fully labeled U-Net model.

\begin{figure}[!t]
    \centering
    \includegraphics[width=0.87\linewidth]{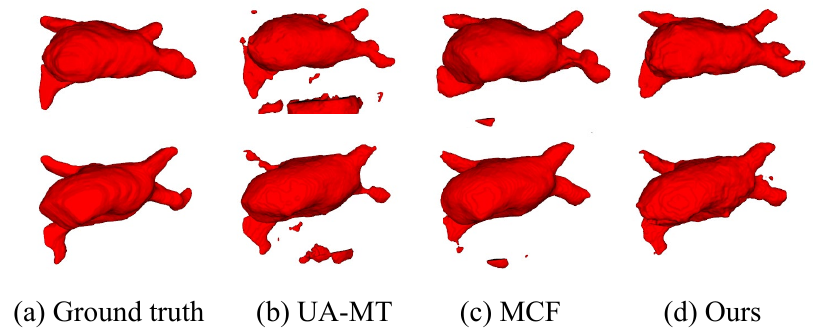}
    \caption{3D segmentation visualization on LA.}
    \label{fig:LA}
    \vspace{-3mm}
\end{figure}

\begin{figure}[!t]
    \centering
    \includegraphics[width=0.75\linewidth]{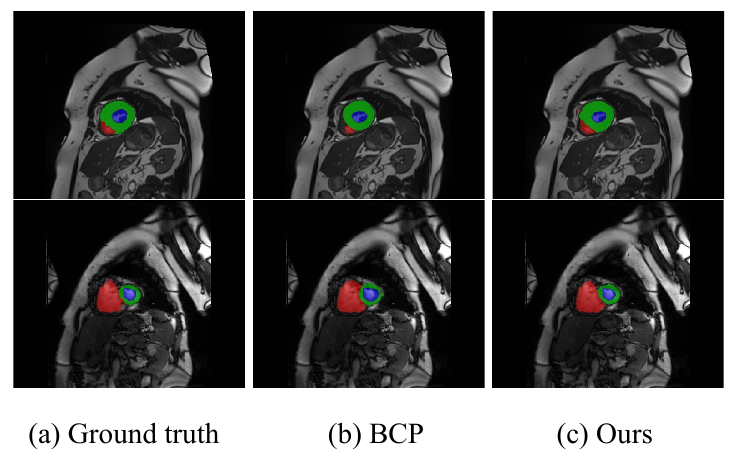}
    \caption{2D segmentation visualization on ACDC.}
    \label{fig:ACDC}
    \vspace{-3mm}
\end{figure}

\noindent\textbf{Qualitative Results.}\ The predicted segmentation results of several examples on the two datasets under 10\% labeling are visualized in Figs.~\ref{fig:LA} and~\ref{fig:ACDC}. Our method produces more precise segmentation than other methods on both 2D and 3D cases.

\begin{table}[!t]
    \caption{Comparison of segmentation quality on ACDC.}
    \label{tab:ACDC}
    \centering
    \sisetup{
		group-separator={,}, 
		group-four-digits=true, 
		table-align-text-post=false,
		detect-weight=true, 
		detect-family=true,
        mode=text
	}
        \begin{tabular}{r | l@{\hspace{1mm}} c | c@{\hspace{1mm}} c@{\hspace{1mm}} S[table-format=2.2]@{\hspace{1mm}} S[table-format=2.2]c}
        \toprule
        Method & Labeled & Unlabeled & Dice$\uparrow$ & Jaccard$\uparrow$ & {95HD$\downarrow$} & {ASD$\downarrow$} \\
        \midrule
        \multirow{3}{*}{U-Net\cite{Ronneberger2015U-net}} & 3(5\%) & 0 & 47.83 & 37.01 & 31.16 & 12.62 \\
        & 7(10\%) & 0 & 79.41 & 68.11 & 9.35 & 2.70 \\
        & 70(All) & 0 & 91.44 & 84.59 & 4.30 & 0.99 \\
        \midrule
        UA-MT \cite{ref30}& \multirow{9}*{3(5\%)} &\multirow{9}*{67(95\%)} & 46.04 & 35.97 & 20.08 & 7.75 \\
        SASSNet \cite{ref31}&   &   & 57.77 & 46.14 & 20.05 & 6.06 \\
        DTC \cite{ref32} &   &   & 56.90 & 45.67 & 23.36 & 7.39 \\
        URPC \cite{ref34}&   &   & 55.87 & 44.64 & 13.60 & 3.74 \\
        MC-Net \cite{ref33}&   &   & 62.85 & 52.29 & 7.62 & 2.33 \\
        SS-Net \cite{ref29} &   &   & 65.83 & 55.38 & 6.67 & 2.28 \\
        Co-BioNet \cite{ref35}&   &    & 87.46&77.93& 1.11&1.11\\
        BCP \cite{ref27}&   &   & 87.59& 78.67& 1.90& 0.67\\
        Ours&   &    &\textbf{88.64}&\textbf{80.21}& \hspace{1.7mm}\textbf{1.51}& \hspace{1.7mm}\textbf{0.60}\\
        \midrule
        UA-MT \cite{ref30}& \multirow{9}*{7(10\%)} &\multirow{9}*{63(90\%)}  & 81.65 & 70.64 & 6.88 & 2.02 \\
        SASSNet \cite{ref31}&   &   & 84.50 & 74.34 & 5.42 & 1.86 \\
        DTC \cite{ref32}&   &   &84.29 & 73.92 & 12.81 & 4.01 \\
        URPC \cite{ref34}&   &   &83.10 & 72.41 & 4.84 & 1.53 \\
        MC-Net \cite{ref33}&   &   &86.44 & 77.04 & 5.50 & 1.84 \\
        SS-Net \cite{ref29}&   &   &86.78 & 77.67 & 6.07 & 1.40 \\
        Co-BioNet \cite{ref35}&   &    &88.49&79.76& 3.70&1.14\\
        BCP \cite{ref27}&   &   & 88.84& 80.62 & 3.98 & 1.17 \\
        Ours&   &    &\textbf{89.69} & \textbf{81.88} & \hspace{1.7mm}\textbf{1.65} & \hspace{1.7mm}\textbf{0.56} \\
        \bottomrule
      \end{tabular}
  \vspace{-3mm}
\end{table}

\subsection{Ablation Study}
\begin{figure}[!t]
    \centering
    \includegraphics[width=0.9\linewidth]{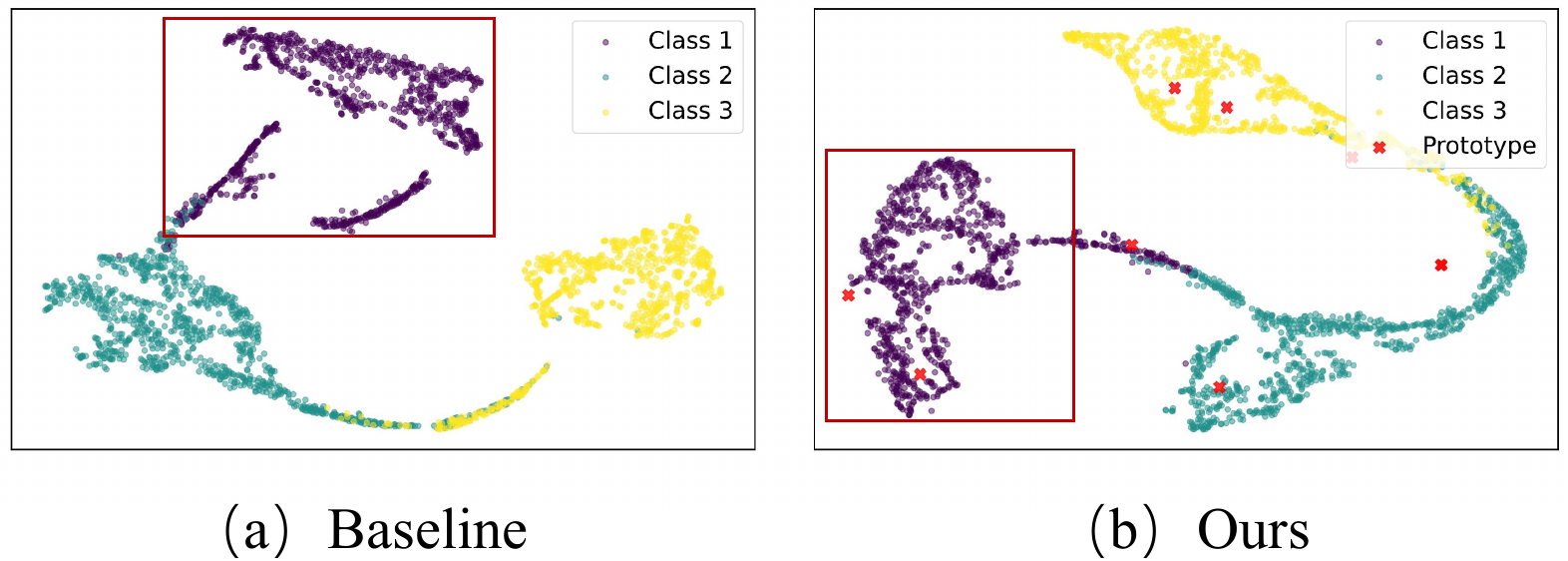}
    \caption{t-SNE visualization on ACDC.}
    \label{fig:t-SNE}
    \vspace{-3mm}
\end{figure}

\noindent\textbf{Key Components.}\ To evaluate the effect of the prototype-based classifier, we implement a baseline model without it. Using 20\% labeled data on the ACDC dataset, t-SNE visualizations (Fig.~\ref{fig:t-SNE}) show that our method achieves more compact embeddings for Class 1. We then train another baseline model, excluding the prototype consistency loss $\mathcal{L}_{cons}^\mathrm{p}$ and the contrastive loss $\mathcal{L}_{cont}$.  Experiments with 5\% labeled data on ACDC (Tab.~\ref{tab:loss}) reveal performance improvements as each loss is added. These results demonstrate the effectiveness of the prototype-based classifier and consistency constraint in modeling intra-class variance.

\begin{table}[!t]
    \caption{Ablation studies of losses on ACDC.} 
    \label{tab:loss}
    \centering
        \begin{tabular}
        {ccc|cccc}
        \toprule
        $\mathcal{L}_{cons}^\mathrm{l}$ & $\mathcal{L}_{cons}^\mathrm{p}$ & $\mathcal{L}_{cont}$  & Dice$\uparrow$ & Jaccard$\uparrow$ & 95HD$\downarrow$ & ASD$\downarrow$ \\
        \midrule
         \checkmark & & &87.18 &78.02&3.73 & 1.07  \\
        \checkmark & \checkmark &  & 88.05 & 79.26 &2.73 &0.80\\
         \checkmark & \checkmark & \checkmark &\textbf{88.64}&\textbf{80.21}&\textbf{1.51}&\textbf{0.60} \\
        \bottomrule
        \end{tabular}
    \vspace{-3mm}
\end{table}

\begin{table}[!t]
    \caption{Ablation studies of prototype update on LA.}
    \label{tab:update}
    \centering
    \begin{tabular}{cc|cccc}
    \toprule
    Labeled & Unabeled   & Dice$\uparrow$ & Jaccard$\uparrow$ & 95HD$\downarrow$ & ASD$\downarrow$ \\
    \midrule
     & &89.55 & 81.25 & 6.83 &1.73\\
     \checkmark&  &  89.76 & 81.53 &6.79&1.80\\
     & \checkmark& 89.57 & 81.25 & 7.10 &1.79\\
     \checkmark & \checkmark & \textbf{90.01}& \textbf{81.94}& \textbf{6.51}& \textbf{1.74} \\
    \bottomrule
    \vspace{-3mm}
\end{tabular}

\end{table}

\begin{table}[!t]
    \caption{Ablation studies of prototype quantity on ACDC.} 
    \label{tab:number}
    \centering
    \begin{tabular}{c|cccc}
    \toprule
    $K$ & Dice$\uparrow$ & Jaccard$\uparrow$ & 95HD$\downarrow$ & ASD$\downarrow$ \\
    \midrule
    1 & 87.10 & 77.93& 3.32&1.05 \\
    2 & 87.52  &78.45&2.59&0.87 \\
    3  &\textbf{88.64}&\textbf{80.21}&\textbf{1.51}&\textbf{0.60}\\  
    4 & 88.11 & 79.43&2.87 &0.83 \\
    5 & 87.98  &79.18 &3.00 &0.76 \\
    \bottomrule
    \end{tabular}
    \vspace{-3mm}
\end{table}
\noindent\textbf{Prototype Update.}\ In the proposed prototype update strategy, both labeled and unlabeled voxels are used. To assess its effect on the segmentation quality, we compare three alternative approaches: no update, using only labeled voxels, and using only unlabeled voxels. Experiments with 10\% labeled data on the LA dataset show that all three strategies degrade performance (Tab.~\ref{tab:update}). This demonstrates that our update strategy builds representative prototypes in each category, and can adapt to potential shifts in data distribution.

\noindent\textbf{Prototype Quantity.}\ Table~\ref{tab:number} shows the performance of our method with different prototype numbers. Experiments with 5\% labeled data on ACDC reveal that when $K = 1$, the prototype degrades into the mean embedding of each class. Performance improves as $K$ increases from 1 to 3, but decreases when $K$ exceeds 3. We speculate that too many sub-centers per class disrupt the compactness. Thus, to balance segmentation accuracy and computational cost, we set $K = 3$.

\section{Conclusion}
We propose a Multi-Prototype-based Embedding Refinement method for semi-supervised medical image segmentation. By leveraging a multi-prototype classification strategy, our method captures intra-class voxel variations and reframes segmentation through structural relationships between voxel embeddings. Furthermore, a consistency constraint integrates different classification granularities. Experiments validate the effectiveness of our approach.


\bibliographystyle{IEEEtran}
\bibliography{IEEEabrv,my}

\end{document}